\documentclass[english]{revtex4-1}
\usepackage[T1]{fontenc}
\usepackage[latin9]{inputenc}
\usepackage{mathrsfs}
\usepackage{amsmath}
\usepackage{amssymb}
\usepackage{graphicx}
\usepackage{esint}

\makeatletter
\usepackage{mathrsfs}
\usepackage{pdfpages}

\DeclareMathOperator*{\argmax}{argmax}

\usepackage{babel}

\makeatother

\usepackage{babel}
\begin{document}



# Subdominant Dense Clusters Allow for Simple Learning and High Computational Performance in Neural Networks with Discrete Synapses

Carlo Baldassi,[1,2,*] Alessandro Ingrosso,[1,2] Carlo Lucibello,[1,2] Luca Saglietti,[1,2] and Riccardo Zecchina[1,2,3]

[1]*Politecnico di Torino, Corso Duca degli Abruzzi 24, I-10129 Torino, Italy*
[2]*Human Genetics Foundation-Torino, Via Nizza 52, I-10126 Torino, Italy*
[3]*Collegio Carlo Alberto, Via Real Collegio 30, I-10024 Moncalieri, Italy*

We show that discrete synaptic weights can be efficiently used for learning in large scale neural systems, and lead to unanticipated computational performance. We focus on the representative case of learning random patterns with binary synapses in single layer networks. The standard statistical analysis shows that this problem is exponentially dominated by isolated solutions that are extremely hard to find algorithmically. Here, we introduce a novel method that allows us to find analytical evidence for the existence of subdominant and extremely dense regions of solutions. Numerical experiments confirm these findings. We also show that the dense regions are surprisingly accessible by simple learning protocols, and that these synaptic configurations are robust to perturbations and generalize better than typical solutions. These outcomes extend to synapses with multiple states and to deeper neural architectures. The large deviation measure also suggests how to design novel algorithmic schemes for optimization based on *local entropy* maximization.

In the past decades, various methods borrowed from statistical physics have been quite successful in studying the basic properties of neural-like systems [1]. A well known general result is that, as is the case for other optimization problems, training neural networks is qualitatively different if the variables—the synaptic weights—are constrained to take discrete values. Yet, the standard equilibrium analysis, which suggests that the solutions to the constrained problem would be inaccessible, is at odds with some recent heuristic algorithmic advances [2–5], which demonstrate that simple effective protocols may be devised at least in some simple scenarios. Therefore, it is conceptually important to understand the underlying reason for this discrepancy, which may be relevant for larger classes of problems.

Furthermore, the modulation of synaptic efficacy is the elementary computational step for biological information storage and for large scale machine learning architectures and neuromorphic devices. While most models assume that synapses are continuous, it is extremely important to understand the practical implications of constraining the synaptic states. Biological considerations and recent experimental evidence [6,7] suggest that synaptic efficacies store a few bits each (between 1 and 5). Machine learning applications (especially hardware implementations) could benefit from using simpler synaptic models and update protocols, a research direction that is currently hampered by the difficulty of devising effective learning protocols.

The learning problem in neural networks with constrained synapses, even in its simplest formulation—the perceptron with $N$ binary synapses—is known to be intractable in the worst case [8]. In the typical case, its equilibrium description is dominated in the large $N$ limit by an exponential number (in $N$) of local minima [9–12], which easily trap standard search strategies based on free energy minimization, e.g. Monte Carlo algorithms [13,14] (a situation typical of spin glass phases, which is common to many hard random optimization problems [15–17]); moreover, the optimal synaptic configurations are typically geometrically isolated (i.e., they have mutual Hamming distances of order $N$), and thus are even harder to find for local search strategies [14].

Here, however, we show that the standard analysis does not capture the properties that are relevant for effective learning strategies. We introduce a novel large deviation analysis that reveals the existence of a different class of solutions, clustered in dense regions. These solutions have radically different properties from the dominating ones, and they are accessible to simple learning protocols. Numerical experiments support the results of this analysis, and show that the same picture also holds for complex neural architectures trained on real-world benchmarks. The analysis generalizes to the case of multilevel synapses.

In a more general sense, these findings highlight the key role that subdominant states have in understanding the practically relevant properties of a prototypical complex system. We have no reason to believe that this scenario is specific to this particular family of problems; in fact, we also show that an optimization strategy inspired by our

analysis is also effective on the random $K$-satisfiability ($K$-SAT) problem.

*The model.*—The single layer binary neural network (perceptron) maps vectors of $N$ inputs $\xi \in \{-1, 1\}^N$ to binary outputs as $\tau(W, \xi) = \text{sgn}(W \cdot \xi)$, where $W \in \{-1, 1\}^N$ is the vector of synaptic weights. Given $\alpha N$ input patterns $\xi^\mu$ with $\mu \in \{1, \ldots, \alpha N\}$ and their corresponding desired outputs $\sigma^\mu \in \{-1, 1\}^{\alpha N}$, and defining $\mathbb{X}_\xi(W) = \prod_{\mu=1}^{\alpha N} \Theta(\sigma^\mu \tau(W, \xi^\mu))$, where $\Theta(x)$ is the Heaviside step function, the learning problem is that of finding $W$ such that $\tau(W, \xi^\mu) = \sigma^\mu$ for all $\mu$, i.e., such that $\mathbb{X}_\xi(W) = 1$. The entries $\xi_i^\mu$ are independent and identically distributed (i.i.d.) unbiased random variables. There are two main scenarios of interest for the distribution of the desired outputs $\sigma^\mu$: (1) the *classification* case, in which they are i.i.d. random variables, and (2) the *generalization* (or *teacher-student*) scenario, in which they are provided by a "teacher" device, i.e., another perceptron with synaptic weights $W^T$. In the classification scenario, the typical problem has a solution with probability 1 in the limit of large $N$ up to $\alpha_c = 0.833$ [9], after which the probability of finding a solution drops to zero. $\alpha_c$ is called the *capacity*; we also use this term for the maximum value of $\alpha$ for which a solution can be found by a specific algorithm. In the teacher-student scenario, the problem has exponentially many solutions up to $\alpha_{TS} = 1.245$, after which there is a first-order transition and only one solution is possible: the teacher itself [1,10]. One additional quantity of interest in this scenario is the generalization error rate $p_e = (1/\pi) \arccos[(1/N) W \cdot W^T]$, which is the probability that $\tau(W, \xi^\star) = \tau(W^T, \xi^\star)$ when $\xi^\star$ is a previously unseen input.

The standard zero-temperature equilibrium analysis of this model is based on a probability measure defined by the partition function $Z_{eq} = \sum_{\{W\}} \mathbb{X}_\xi(W)$; the typical case is described by taking the quenched average $\langle \log(Z_{eq}) \rangle_\xi$ over the realizations of the patterns.

*Effective learning algorithms.*—Only a handful of heuristic algorithms are currently believed—based on numerical evidence—to be able to solve the classification problem and achieve a nonzero capacity in the limit of large $N$ in a subexponential running time: reinforced Belief Propagation (BP) [2], reinforced Max-Sum [3], SBPI [4], and CP + R [5] (a brief description of each is provided in the Supplemental Material [18]). In the classification case, they achieve capacities between $\alpha \simeq 0.69$ and $\alpha \simeq 0.75$. They all share the property of being local and distributed, and have typical solving times that scale almost linearly with the size of the input. SBPI and CP + R additionally have extremely simple requirements (only employing finite discrete quantities and simple, local, and online update schemes), making them appealing for practical purposes and reasonably plausible candidates for biological implementations. A qualitatively similar scenario holds in the generalization case, where all these algorithms perform well except in a finite window $1 \lesssim \alpha \lesssim 1.5$ around $\alpha_{TS}$.

These results are not captured by the standard spin glass theory; in particular, the effectiveness of the utterly simplified algorithms SBPI and CP + R is in striking contrast with a glassy energy landscape in which solutions are isolated.

*Numerical experiments.*—We investigated this issue numerically, and found evidence that, in fact, the solutions found by the algorithms are typically not isolated; rather, they belong (with high probability at large $N$) to large connected clusters of solutions. More precisely: (1) from a given solution $\tilde{W}$, a random walk process over neighboring configurations in the space of solutions can reach distances of order $N$ from the starting point; (2) the number of solutions at a distance of order $N$ from $\tilde{W}$ grows exponentially with $N$ (this can be estimated from the analysis of the recurrence relations on the average growth factor of the number of solutions at varying distances, and using the random walk processes for sampling the local properties relevant to those relations).

Furthermore, we used the standard BP method on single problem instances to estimate the entropy of the solutions at varying distance (controlled via a Franz-Parisi potential [21]) from a reference solution $\tilde{W}$ obtained from a heuristic solver, and found that the results do not match the predictions of the equilibrium analysis [14], see Fig. 1.

*Teacher-student case.*—We also extended the equilibrium analysis [14] to the teacher-student scenario, and found that: (1) typical solutions are isolated for all values of $\alpha$ even when adding a nonzero stability constraint, as in the classification case; (2) the teacher device is isolated and indistinguishable from all other typical solutions except for the generalization error; and (3) the results of estimates obtained from BP are consistent with the analytical calculation when using the teacher as a reference point, but not when using a solution provided by a heuristic solver (see the inset in Fig. 1). Finally, the generalization error for solutions found algorithmically is lower than what would be expected for a typical solution (see Fig. 3).

*Large deviation analysis.*—These results indicate that calculations performed at thermodynamic equilibrium are effectively blind to the solutions found by the heuristic algorithms. Traditionally, in the context of replica theory, similar situations have been addressed by looking for subdominant states [22]. However, this is insufficient in the present case.

A different analytical tool is thus needed for obtaining a description of this regime, which—according to the numerical evidence—is characterized by regions with a high density of solutions. Clearly, the statistical weight of the individual solutions must be modified, by favoring the ones that are surrounded by a large number of other solutions. Therefore, we studied the following large-deviation free energy density function:

$$\mathcal{F}(d, y) = -\frac{1}{Ny} \log \left( \sum_{\{\tilde{W}\}} \mathbb{X}_\xi(\tilde{W}) \mathcal{N}(\tilde{W}, d)^y \right) \quad (1)$$

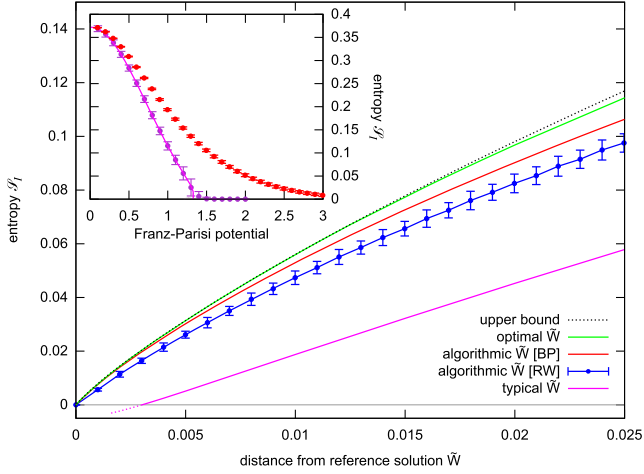

FIG. 1 (color online). Numerical evidence of the existence of clusters of solutions. Entropy at a given distance from a reference solution $\tilde{W}$, in the classification case at $\alpha = 0.4$. From bottom to top: (magenta) theoretical prediction for a typical $\tilde{W}$; (blue) numerical estimate based on a random walks on connected solutions starting from one provided by SBPI, with $N = 1001$; (red) estimate from belief propagation using a solution from SBPI, with $N = 10001$; (green) theoretical curve for the optimal $\tilde{W}$ as computed from Eq. (1); and (dotted black) upper bound ($\alpha = 0$ case, all configurations are solutions). The random-walk points underestimate the number of solutions since they only consider single-flip-connected clusters; the BP curve is lower than the optimal curve because in the latter $\tilde{W}$ is optimized as a function of the distance, while in the former it is fixed. Inset: comparison between a typical solution and one found with SBPI, in the teacher-student case at $\alpha = 0.5$ with $N = 1001$. Larger potentials correspond to smaller distances. Top points (red): SBPI reference solution, with the entropy computed by BP; bottom curve (magenta): theoretical prediction for a typical solution; bottom points (purple): BP results using the teacher as reference.

where $\mathcal{N}(\tilde{W}, d) = \sum_{\{W\}} \mathbb{X}_\xi(W) \delta(W \cdot \tilde{W}, N(1-2d))$ counts the number of solutions $W$ at normalized Hamming distance $d$ from a reference solution $\tilde{W}$ ($\delta$ is the Kronecker delta symbol) and $y$ has the role of an inverse temperature. This free energy describes a system in which each configuration $\tilde{W}$ is constrained to be a solution, and has a formal energy density $\mathcal{E}(\tilde{W}) = -(1/N) \log \mathcal{N}(\tilde{W}, d)$ that favors configurations surrounded by an exponential number of other solutions, with $y$ controlling the amount of reweighting. The regions of highest local density are then described in the regime of large $y$ and small $d$.

The relevant quantities are computed through the usual statistical physics tools; of particular importance is the entropy density of the surrounding solutions, the *local entropy*:

$$\mathcal{S}_I(d, y) = -\langle \mathcal{E}(\tilde{W}) \rangle_{\xi, \tilde{W}} = \frac{1}{N} \langle \log \mathcal{N}(\tilde{W}, d) \rangle_{\xi, \tilde{W}}, \quad (2)$$

which is simply given by $\mathcal{S}_I(d, y) = \partial_y(y \mathcal{F}(d, y))$. The signature for the existence of a dense and exponentially large cluster of solutions is that $\mathcal{S}_I(d, y) > 0$ in a neighborhood of $d = 0$. Another important quantity is the *external entropy*, i.e., the entropy of the reference solutions $\mathcal{S}_E(d, y) = -y[\mathcal{F}(d, y) + \mathcal{S}_I(d, y)]$, which must also be non-negative.

The special case $y = 1$ is essentially equivalent to the computation of Ref. [11]; $\mathcal{S}_I(d, y)$ reduces to the computation à la Franz-Parisi of [14] in the limit $y \to 0$.

We computed Eq. (1) by the replica method in the replica-symmetric (RS) ansatz, resulting in an expression involving 13 order parameters to be determined by the saddle point method. The analytical expressions and the details of the computation are reported in the Supplemental Material [18]. It turns out that, for all values of $\alpha$ and $d$, there is a value of $y$ beyond which $\mathcal{S}_E(d, y) < 0$, which is unphysical and signals a problem with the RS assumption. Therefore, we sought the value $y^\star = y^\star(\alpha, d)$ at which $\mathcal{S}_E(d, y^\star) = 0$, i.e., the highest value of $y$ for which the RS analytical results are consistent. In the following, we thus drop the $y$ dependency.

The solution to the system of equations stemming from the RS saddle point produces qualitatively very similar results for both the classification (with $\alpha < \alpha_c$) and the generalization (with $\alpha < \alpha_{TS}$) case. It displays a number of noteworthy properties (Fig. 2):

(1) For all $\alpha < \alpha_c$, there is a neighborhood of $d = 0$ where $\mathcal{S}_I(d) > 0$, implying the existence of extensive

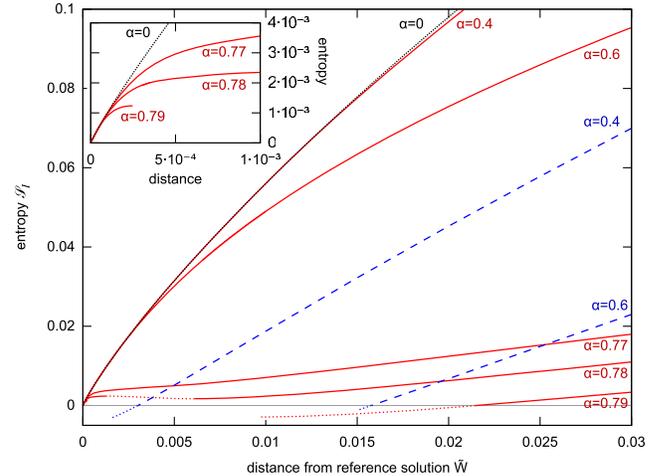

FIG. 2 (color online). Large deviation analysis. Local entropy curves at varying distance $d$ from the reference solution $\tilde{W}$ for various $\alpha$ (classification case). Black dotted curve, $\alpha = 0$ case (upper bound). Red solid curves, RS results from Eq. (1) (optimal $\tilde{W}$). Up to $\alpha = 0.77$, the curves are monotonic. At $\alpha = 0.78$, a region incorrectly described within the RS ansatz appears (dotted; geometric bounds are violated at the boundaries of the part of the curve with negative derivative). At $\alpha = 0.79$, the solution is discontinuous (a gap appears in the curve), and parts of the curve have negative entropy (dotted). Blue dashed curves, equilibrium analysis (typical $\tilde{W}$) [14] (dotted parts are unphysical): the curves are never positive in a neighborhood of $d = 0$. Inset: enlargement of the region around $d = 0$ (notice the solution for $\alpha = 0.79$, followed by a gap).

clusters of solutions. Furthermore, for all $\alpha$, the curves for $\mathcal{S}_I(d)$ are all approximately equal around $d = 0$; in particular, they all approximate the case for $\alpha = 0$ where all points are solutions. This implies that the clusters of solutions are extremely dense at their core. This is our chief result. The size of this dense region shrinks with $\alpha$ and vanishes at $\alpha_c$.

(2) For large distances, as expected, $\mathcal{S}_I(d)$ collapses with a second-order transition onto the equilibrium entropy; i.e., this regime is dominated by the typical solutions.

(3) Up to a certain $\alpha_U$ (where $\alpha_U \simeq 0.77$ in the classification case and $\alpha_U \simeq 1.1$ in the generalization case), the $\mathcal{S}_I(d)$ curves are monotonic in $d$. Beyond $\alpha_U$, there is a transition in which there appear regions of $d$ (dotted in Fig. 2) that are not correctly described by the RS ansatz (since geometric bounds are violated; see the discussion in the Supplemental Material for details [18]), and must be described at a higher level of replica symmetry breaking (RSB). We speculate that this transition signals a change in the structure of the space of solutions: for $\alpha < \alpha_U$, the densest cores of solutions are immersed in a huge connected structure; for $\alpha > \alpha_U$, this structure fractures and the dense cores become isolated and hard to find.

(4) In the teacher-student scenario, the generalization properties of the optimal reference solutions $\tilde{W}$ are generally much better than those of typical solutions. This is clearly shown in Fig. 3, where we also show that the curve for small $d$ is in striking agreement with that produced using solutions obtained from the SBPI algorithm. The generalization error decreases monotonically when increasing $d$, and it saturates to a plateau when $\mathcal{S}_I(d)$ becomes equal to the entropy of the typical solutions [see point (2) above].

We expect this qualitative and quantitative picture, especially for $\alpha \lesssim \alpha_U$, to be quite robust. First, these results are convincingly supported by our numerical findings, where available. Furthermore, a slightly simplified model analyzed at a higher level of RSB and at $y \to \infty$ [see Eq. (3) below] yields almost indistinguishable results.

The analytical computations are straightforwardly generalized to the case of multilevel synapses and sparse patterns, and the results are qualitatively identical [23].

*Multilayer network.*—These theoretical results seem to extend to more complex architectures and nonrandom learning problems. We observed this by heuristically extending the CP + R algorithm to multilayer classifiers with $L$ possible output labels, and training these networks on the MNIST database benchmark [24], which consists of $7 \times 10^4$ grayscale images of hand-written digits ($L = 10$). A description of the architecture and of the learning algorithm is provided in the Supplemental Material [18].

We observed that it is indeed very easy to achieve perfect learning on the whole training data set, and that very good generalization errors can be reached (e.g., 1.25% with order $10^7$ synapses) despite the binary nature of the synapses and the fact that we did not specialize the architecture for this particular data set. Moreover, we did not observe any overfitting: the generalization error does not degrade by reaching zero training error, or by using larger networks.

As for the perceptron, we performed a random-walk process in the space of solutions, with similar results: the simplified algorithm reaches a solution that is part of a dense, large connected cluster, and the generalization properties of the starting solution are better than those of solutions found in later stages of the random walk (see Fig. 1B in the Supplemental Material [18]).

*Optimization.*—We also studied a variant of the free energy (1) without the constraint on $\tilde{W}$:

$$\mathcal{F}_U(d, y) = -\frac{1}{Ny} \log \left( \sum_{\{\tilde{W}\}} \mathcal{N}(\tilde{W}, d)^y \right). \qquad (3)$$

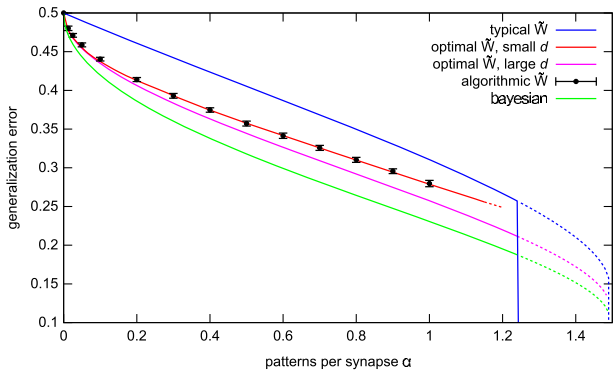

FIG. 3 (color online). Generalization error (teacher-student scenario). From top to bottom: (blue) typical solution, (red) optimal $\tilde{W}$ from Eq. (1) at small $d$ (we used $d = 0.025$ for numerical reasons and since the curve is not sensitive to the precise value of $d$ in this regime; this solution disappears after $\alpha \simeq 1.2$), (black points) solutions from SBPI at $N = 10001$, 100 samples per point, (magenta) optimal $\tilde{W}$ from Eq. (1) at the value of $d$ for which $\mathcal{S}_I$ is maximum (i.e., it equals the equilibrium entropy), and (green) Bayesian case: error from the average over all solutions. At $\alpha_{TS} = 1.245$ there is the first-order transition to perfect learning; between $\alpha_{TS}$ and $\alpha = 1.5$ there is a metastable regime; the dashed parts of the curves correspond to unphysical solutions of the RS equations with negative entropy.

The analysis in this case requires at least an additional step of RSB, and will be presented in detail in a follow-up work [25]. Still, the results are very close to those reported for the constrained scenario; furthermore, the probability that the $\tilde{W}$ in this system are a solution tends exponentially to 1 with $d \to 0$, despite the removal of the explicit constraint. This suggests that we can algorithmically exploit $\mathcal{F}_U(d, y)$ to efficiently sample ground states of the system, and that such a strategy could be applied to different optimization problems as well.

As the most straightforward proof of concept in this direction we have developed a Monte Carlo Markov Chain

algorithm, using the local entropy $\mathcal{E}(\tilde{W})$ as an objective function. We call such a procedure the Entropy-driven Monte Carlo (EdMC) procedure [25]. $\tilde{W}$ is initialized at random; at each step $\mathcal{E}(\tilde{W})$ is computed efficiently by the BP algorithm, which is expected to give good results at small $d$ in dense regions; random local updates of $\tilde{W}$ are accepted or rejected using a standard Metropolis rule at fixed temperature $y^{-1}$. In addition to the binary perceptron problem, we applied the algorithm to another (radically different) problem in which standard simulated annealing (SA) methods are known to fail, namely, the famous random $K$-SAT problem. As expected from our analysis, for the perceptron learning problem EdMC does not suffer from trapping in local minima even at $y = \infty$, up to at least $\alpha \simeq 0.65$, with a running time that scales almost linearly with the size of the problem (whereas SA solving time diverges exponentially). For the random $K$-SAT problem we explored different regimes, and in particular one (in 4-SAT) where SA method is known to fail: in all cases, EdMC succeeds in an almost linear number of steps, outperforming SA method. Preliminary quantitative data are given in the Supplemental Material [18].



\title{Subdominant dense clusters allow for simple learning and high computational
performance in neural networks with discrete synapses\\
SUPPLEMENTAL MATERIAL}

\author{Carlo Baldassi}

\author{Alessandro Ingrosso}

\author{Carlo Lucibello}

\author{Luca Saglietti}

\affiliation{Politecnico di Torino, Corso Duca degli Abruzzi 24, I-10129 Torino,
Italy}

\affiliation{Human Genetics Foundation-Torino, Via Nizza 52, I-10126 Torino, Italy}

\author{Riccardo Zecchina}

\affiliation{Politecnico di Torino, Corso Duca degli Abruzzi 24, I-10129 Torino,
Italy}

\affiliation{Human Genetics Foundation-Torino, Via Nizza 52, I-10126 Torino, Italy}

\affiliation{Collegio Carlo Alberto, Via Real Collegio 30, I-10024 Moncalieri,
Italy}

\maketitle

\subsection{Brief description of the heuristic algorithms}

As mentioned in the main text, there are 4 algorithms which are currently
known to be able to solve the classification problem for large $N$
in a sub-exponential running time: reinforced Belief Propagation (R-BP)
\cite{braunstein-zecchina}, reinforced Max-Sum (R-MS) \cite{baldassi-braunstein-maxsum},
SBPI \cite{baldassi-et-all-pnas} and CP+R \cite{baldassi-2009}.
Here, we provide a brief summary of their characteristics.

The R-BP algorithm is a variant of the standard Belief Propagation
(BP) algorithm. BP is a cavity method which can be used to compute
the equilibrium properties (e.g.~marginal probability distributions
of single variables, entropy, etc.) at a given temperature for a particular
instance of a problem described in terms of a factor graph. It is
based on the Bethe-Peierls approximation, and it is known to give
exact results under some circumstances in the limit of large $N$;
in particular, for the case of the binary perceptron with random inputs,
it is believed that this is the case below $\alpha_{c}$. The BP algorithm
can be turned into a heuristic solver by adding a reinforcement term:
this is a time-dependent external field which tends to progressively
polarize the probability distributions on a particular configuration,
based on the approximate marginals computed at preceding steps of
the iteration of the BP equations. The reinforcement can thus be seen
as a ``soft decimation'' process, in which the variables are progressively
and collectively fixed until they collapse onto a single configuration.
This method seems to have an algorithmic capacity of at least $\alpha\simeq0.74$.

The R-MS algorithm is analogous to the R-BP algorithm, using Max-Sum
(MS) as the underlying algorithm rather then BP. The MS algorithm
can be derived as a particular zero-temperature limit of the BP equations,
or it can be seen as a heuristic extension of the dynamic programming
approach to loopy graphs. The reinforcement term acts in the same
way as previously described for R-BP. The resulting characteristics
of R-MS are very similar to those of BP; extensive numerical tests
give a capacity of about $\alpha\simeq0.75$.

The SBPI algorithm was derived as a crude simplification of the R-BP
algorithm, the underlying idea being that of stripping R-BP of all
features which would be completely unrealistic in a biological context.
This resulted in an on-line algorithm, in which patterns are presented
one at a time, and in which only information locally available to
the synapses is used in the synaptic update rule. Furthermore, the
algorithm only uses a finite number of discrete internal states in
each synapse, and is remarkably robust to noise and degradation. Rather
surprisingly, despite the drastic simplifications, the critical capacity
of this algorithm is only slightly reduced with respect to the original
R-BP algorithm, and was measured at about $\alpha\simeq0.69$.

The CP+R algorithm was derived as a further simplification of the
SBPI algorithm. It is equivalent to the former in the context of the
on-line generalization task, but requires some minor modifications
in the classification context. Its main difference from SBPI is that
it substitutes an update rule which was triggered by near-threshold
events in SBPI with a generalized, stochastic, unsupervised synaptic
reinforcement process (the rate of application of this mechanism needs
to be calibrated for optimal results). Note that the kind of reinforcement
mentioned here is rather different from the reinforcement term of
R-BP or R-MS. The capacity of the CP+R algorithm can be made equal
to that of SBPI, $\alpha\simeq0.69$.

\subsection{Large deviation analysis}

We computed the action $\phi=-y\mathscr{F}$ corresponding to the
free energy $\mathscr{F}$ of eq.~$\left(1\right)$ of the main text by the replica
method, in the so called replica-symmetric (RS) Ansatz. The resulting
expression, in the generalization case, is:
\begin{eqnarray}
\phi\left(S,y\right) & = & -\frac{1}{2}\left(1-\tilde{q}\right)\hat{\tilde{q}}-\frac{y}{2}\left(1-q_{1}\right)\hat{q}_{1}-\frac{y^{2}}{2}\left(q_{1}\hat{q}_{1}-q_{0}\hat{q}_{0}\right)+y\tilde{S}\hat{\tilde{S}}-yS\hat{S}-\tilde{R}\hat{\tilde{R}}-yR\hat{R}+\mathcal{G}_{S}+\alpha\mathcal{G}_{E}\label{eq:action}
\end{eqnarray}
where we used the overlap $S=1-2d$ as a control parameter instead
of $d$, and
\begin{eqnarray*}
\mathcal{G}_{S} & = & \int D\tilde{z}\int Dz_{0}\,\log\sum_{\tilde{W}=\pm1}\exp\left(\tilde{W}\tilde{A}\left(\tilde{z},z_{0}\right)\right)\int Dz_{1}\left(2\cosh\left(A\left(z_{0},z_{1},\tilde{W}\right)\right)\right)^{y}
\end{eqnarray*}
\[
\mathcal{G}_{E}=2\int D\tilde{z}\int Dz_{0}\,H\left(\eta\left(\tilde{z},z_{0}\right)\right)\log\int Dz_{1}H\left(\tilde{C}\left(\tilde{z},z_{0},z_{1}\right)\right)H\left(C\left(z_{0},z_{1}\right)\right)^{y}
\]
\begin{eqnarray*}
\tilde{A}\left(\tilde{z},z_{0}\right) & = & \tilde{z}\sqrt{\hat{\tilde{q}}-\frac{\hat{\tilde{S}}^{2}}{\hat{q}_{0}}}+z_{0}\frac{\hat{\tilde{S}}}{\sqrt{\hat{q}_{0}}}+\hat{\tilde{R}}\\
A\left(z_{0},z_{1},\tilde{W}\right) & = & z_{1}\sqrt{\hat{q}_{1}-\hat{q}_{0}}+z_{0}\sqrt{\hat{q}_{0}}+\hat{R}+\tilde{W}\left(\hat{S}-\hat{\tilde{S}}\right)\\
\eta\left(\tilde{z},z_{0}\right) & = & \frac{wRz_{0}+\left(q_{0}\tilde{R}-R\tilde{S}\right)\tilde{z}}{\sqrt{q_{0}}\sqrt{w^{2}-\left(\tilde{q}R^{2}+q_{0}\tilde{R}^{2}-2R\tilde{R}\tilde{S}\right)}}\\
w & = & \sqrt{\tilde{q}q_{0}-\tilde{S}^{2}}\\
C\left(z_{0},z_{1}\right) & = & \frac{z_{1}\sqrt{q_{1}-q_{0}}+z_{0}\sqrt{q_{0}}}{\sqrt{1-q_{1}}}\\
\tilde{C}\left(\tilde{z},z_{0},z_{1}\right) & = & \frac{\left(S-\tilde{S}\right)z_{1}+\sqrt{\frac{q_{1}-q_{0}}{q_{0}}}\left(\tilde{S}z_{0}+w\tilde{z}\right)}{\sqrt{\left(q_{1}-q_{0}\right)\left(1-\tilde{q}\right)-\left(S-\tilde{S}\right)^{2}}}
\end{eqnarray*}

The order parameters $\tilde{q}$, $q_{1}$, $q_{0}$, $\tilde{S}$,
$R$, $\tilde{R}$ and their conjugates $(\hat{\tilde{q}}$, $\hat{q}_{1}$
etc. and $\hat{S}$) must be determined from the saddle point equations,
i.e.~by setting to zero the derivative of $\phi\left(S,y\right)$
with respect to each parameter. This yields a system of $13$ coupled
equations, with $\alpha$, $y$ and $S$ as control parameters. We
solved these equations iteratively.

The physical interpretation of the order parameters is as follows
(here, the overlap between two configurations $X$ and $Y$ is defined
as $\frac{1}{N}\left(X\cdot Y\right)$):
\begin{description}
\item [{$\tilde{q}$}] overlap between two different reference solutions
$\tilde{W}$
\item [{$q_{1}$}] overlap between two solutions $W$ referred to the same
$\tilde{W}$
\item [{$q_{0}$}] overlap between two solutions $W$ referred to two different
$\tilde{W}$
\item [{$S$}] overlap between a solution $W$ and its reference solution
$\tilde{W}$
\item [{$\tilde{S}$}] overlap between a solution $W$ and an unrelated
reference solution $\tilde{W}$
\item [{$R$}] overlap between a solution $W$ and the teacher $W^{\mathscr{T}}$
\item [{$\tilde{R}$}] overlap between a reference solution $\tilde{W}$
and the teacher $W^{\mathscr{T}}$
\end{description}
Therefore, $\tilde{R}$ can be used to compute the typical generalization
error of reference solutions $\tilde{W}$, as $\frac{1}{\pi}\arccos\left(\tilde{R}\right)$.
An analogous relation yields the generalization error of the solutions
$W$ as a function of $R$.

It is also worth noting that $\tilde{q}<1$ implies that the number
of reference solutions $\tilde{W}$ is larger than $1$.

By setting to zero the order parameters $R$, $\tilde{R}$ and their
conjugates, and thus reducing the system of equations to the remaining
$9$ saddle point conditions, we obtain the classification scenario.

It can be noted that, although we call this solution replica-symmetric,
the structure is highly reminiscent of a $1$-RSB solution. Indeed,
it can be shown that, if we remove the constraints on the configurations
$\tilde{W}$, and solve for $\tilde{S}=0$ rather then fixing $S$,
we obtain exactly the standard $1$-RSB equations for the perceptron
of \cite{krauth-mezard} at zero temperature, with $y$ taking the
role of the Parisi parameter $m$. However, the $1$-RSB solution of the
standard equations shows no hint of the dense regions which we find
in the present work, even if we relax the requirement $0\le m\le1$
of \cite{krauth-mezard}. This shows that the constraint on the distance
is crucial to explore these sub-dominant regions.

From eq.~(\ref{eq:action}) we can compute the internal and external
entropies, as:
\begin{eqnarray}
\mathscr{S}_{I}\left(S,y\right) & = & \frac{\partial\phi}{\partial y}\left(S,y\right)\\
\mathscr{S}_{E}\left(S,y\right) & = & \phi\left(S,y\right)-y\frac{\partial\phi}{\partial y}\left(S,y\right)
\end{eqnarray}

From the last equation, we define $y^{\star}$ by $\mathscr{S}_{E}\left(S,y^{\star}\right)=0$.
We sought this value numerically for each $\alpha$ and $S$. Therefore,
in all our results, the typical number of reference solutions $\tilde{W}$
was sub-exponential in $N$; however, we found that in all cases $\tilde{q}<1$,
which implies that the solutions $\tilde{W}$ are not unique.

Using the value of the temperature at which the (external) entropy
vanishes is sufficient in this case to derive results which are geometrically
valid across most values of the control parameters $\alpha$ and $S$.
As noted in the main text (see also Fig.~\textbf{$2$} in the main
text), there are two exceptions to this observation, both occurring
at high values of $\alpha$ and in specific regions of the parameter
$S$ ($d$ in the main text). Let us indicate with $\left[S_{L},S_{R}\right]$
these regions, with $0<S_{L}<S_{R}<1$. The most obvious kind of problem
occurs occurs at $\alpha\gtrsim0.79$, where $\mathscr{S}_{I}\left(S,y\right)<0$
for $S\in\left[S_{L},S_{R}\right]$. Another type of transition occurs
between $\alpha\simeq0.77$ and $\alpha\simeq0.79$, where the $\frac{\partial}{\partial S}\mathscr{S}_{I}\left(S,y\right)\ge0$
in $\left[S_{L},S_{R}\right]$. A closer inspection of the order parameters
reveals that, $q_{1}\ge S$ for $S\in\left[S_{L,}S_{R}\right]$ .
The transition points $S_{L}$ and $S_{R}$ at which $q_{1}=S$ are
manifestly unphysical, because in that case any of the solutions $W$
(which are exponential in number, since $\mathscr{S}_{I}>0$) could
play the role of the reference solution $\tilde{W}$, and yet the
number of $\tilde{W}$ should be sub-exponential, because $\mathscr{S}_{E}=0$.
This is a contradiction. We conclude that those regions are inadequately
described within the RS Ansatz as well.

As for the parts of the curves which are outside these problematic
regions, the results obtained under the RS assumption are reasonable,
and in very good agreement with the numerical evidence. In order to
assess whether the RS equations are stable, further steps of RSB would
be needed; unfortunately, this would multiply the number of order
parameters (and thus enlarge the system of equations) and the number
of nested integrals required for each of these equations, which is
computationally too heavy at the present time. Also, we should observe
that the true extremal cases are described only in the limit of $y\to\infty$,
for which the RS solution is inadequate, and thus that our reported
values of $\mathscr{S}_{I}$ are probably a lower bound. Note, however,
that $y^{\star}\to\infty$ both when $S\to1$, i.e.~in the limit
of small distances where the solutions exhibit the highest density,
and at small $S$, i.e.~where the saddle point solution encompasses
the typical equilibrium solutions of the standard analysis and $\mathscr{S}_{I}$
becomes equal to the standard entropy of the equilibrium ground
states.

In conclusion, our results suggest that the general picture is well
described by the RS assumption with the zero external entropy requirement,
and that quantitative adjustments due to further levels of RSB would
likely be small, and limited to the intermediate regions of $S$.

\subsection{Multi-layer network with binary synapses}

\begin{figure}
\includegraphics[width=1\columnwidth]{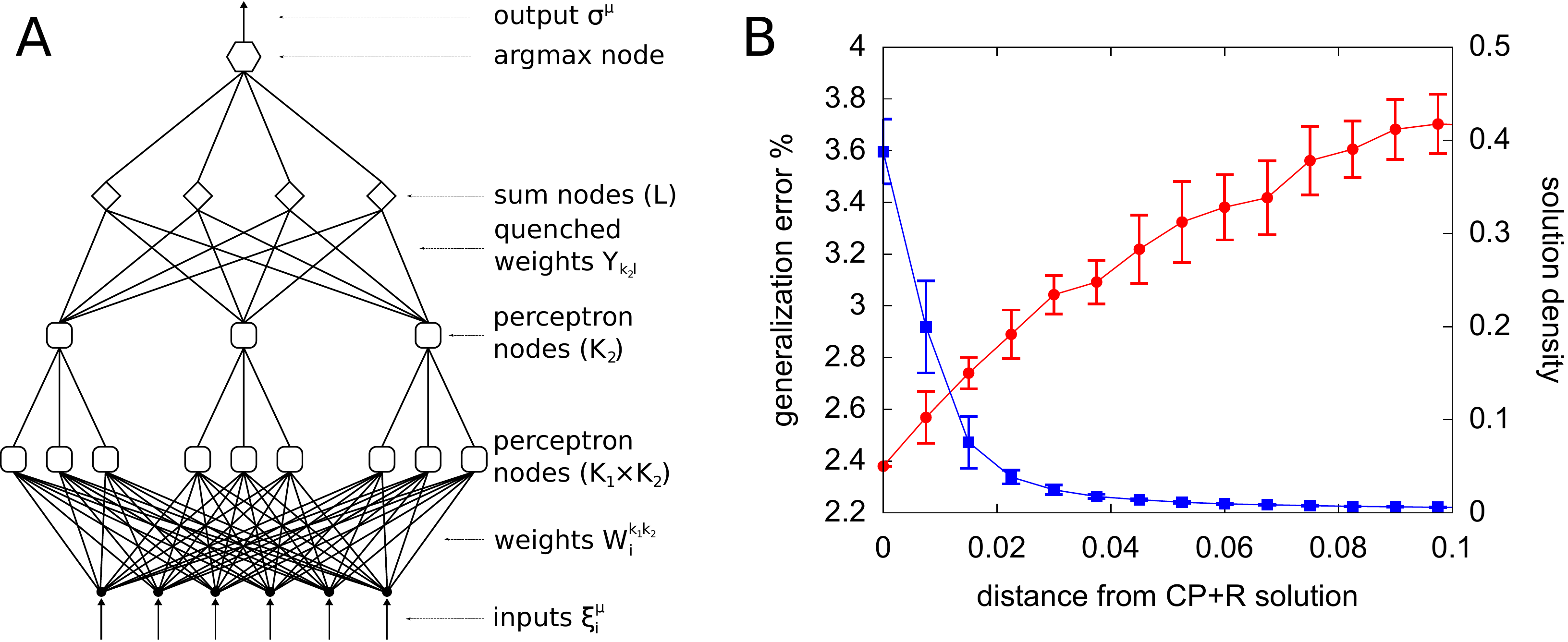}\protect\protect\caption{\label{fig:deep}(Color online) \textbf{Multi-layer tests on MNIST}.
\emph{A}. Network scheme. \emph{B}. results of a random walk over
solutions to the training set (with $K_{1}=11$, $K_{2}=30$, $r=0$),
starting from a solution found by CP+R. Moving away from this solution,
the generalization error (red, circles) increases, and the solution
density (blue, squares) decreases. The same qualitative behavior is
observed with all network sizes, and regardless of preprocessing.}
\end{figure}

We heuristically extended the CP+R algorithm to multi-layer classifiers
with $L$ possible output labels. The architecture we used (Fig.~\ref{fig:deep}A)
consists of an array of $K_{2}$ committee machines, each comprising
$K_{1}$ hidden units, whose outputs are sent to $L$ summation nodes,
and from these to a readout node which performs an $\argmax$ operation.
This network therefore realizes the following map: 
\[
\psi\left(\xi\right)=\argmax_{l\in\left\{ 1,\dots,L\right\} }\left(\sum_{k_{2}=1}^{K_{2}}Y_{k_{2}l}\,\textrm{sign}\left(\sum_{k_{1}=1}^{K_{1}}\tau\left(W^{k_{1}k_{2}},\xi_{i}\right)\right)\!\!\right)
\]
where $Y_{k_{2}l}\in\left\{ -1,1\right\} $ are random quenched binary
weights, and $W^{k_{1}k_{2}}\in\left\{ -1,1\right\} ^{N}$ are the
synaptic weights.

The single-layer CP+R rule consists of two independent processes,
a supervised one and a generalized, unsupervised one (see \cite{baldassi-2009}
for details). For the multi-layer case, we kept the unsupervised process
unaltered, and used a simple scheme to back-propagate the error signals
to the individual perceptron units, as follows: upon presentation
of a pattern $\xi$ whose required output is $\sigma$, in case of
error ($\psi\left(\xi\right)\ne\sigma$), a signal is sent back to
all committee machines which contributed to the error, i.e.~all those
for which $\textrm{sign}\left(\sum_{k_{1}=1}^{K_{1}}\tau\left(W^{k_{1}k_{2}},\xi_{i}\right)\right)\ne Y_{k_{2}\sigma}$.
Each of these in turn propagates back a signal to the hidden unit,
among those which provided the wrong output (i.e.~for which $Y_{k_{2}\sigma}\sum_{i=1}^{N}W_{i}^{k_{1}k_{2}}\,\xi_{i}<0$),
which is the easiest to fix, i.e.~for which $-Y_{k_{2}\sigma}\sum_{i=1}^{N}W_{i}^{k_{1}k_{2}}\,\xi_{i}$
is minimum. Finally, the hidden units receiving the error signal update
their internal state according to the CP+R supervised rule. We also
added a ``robustness'' setting, such that an error signal is emitted
also when $\psi\left(\xi\right)=\sigma$, but the difference between
the maximum and the second maximum in the penultimate layer is smaller
then some threshold $r$.

We tested this network on the MNIST database benchmark \cite{lecun1998gradient},
which consists of $7\cdot10^{4}$ grayscale images of hand-written
digits ($L=10$); of these, $10^{4}$ are reserved for assessing the
generalization performance. The images were subject to standard unsupervised
preprocessing by a Restricted Boltzmann Machine ($N=501$ output nodes)
\cite{hinton2006fast,bengio2013representation}, but this is not essential
for training: the inputs could be used directly, or be simply pre-processed
by random projections, with only minor effects on the performance.
The smallest network which is able to perfectly learn the whole training
dataset had $K_{1}=11$ and $K_{2}=30$, with $r=0$; its generalization
error was about $2.4\%$. Larger networks achieve better generalization
error rates, e.g.~$1.25\%$ with $K_{1}=81$, $K_{2}=200$, $r=120$.

\subsection{Optimization}

\subsubsection*{Perceptron}

Entropy driven Monte Carlo (EdMC) was applied and confronted with
Simulated Annealing (SA) at increasing $N$ for different values of
$\alpha$: the proposed strategy was able to reach a solution, i.e.~a
configuration at zero energy, in a time which scales almost linearly
with $N$, while as expected SA often gets stuck in local minima even
at low loading and with an extremely slow cooling rate.

As an example at $\alpha=0.3$ and $N\in\left\{ 201,401,801,1601\right\} $,
we studied the EdMC behavior over $100$ random instances of the classification
problem, and found that the number of required iterations scales approximately
as $N^{1.2}$.

\subsubsection*{Random K-satisfiability}

For the random K-satisfiability problem we explored two regions of
parameters: 3-SAT in its RS phase, where both EdMC and Simulated Annealing
are expected to succeed, and random 4-SAT in the RSB regime where
SA is known to fail. As for the perceptron problem, we observe a much
faster running time in favor of EdMC in both cases.

For the 4-SAT case, in order to bypass the convergence problems of
BP, it's possible to use temporal averages to approximate the local
entropy. This technical problem should however be overcome by computing
the entropy at the 1-RSB level, which is beyond the scope of this
preliminary study.

As an example, for values of $N$ ranging in $\left\{ 100,500,1000,5000,10000\right\} $
and a number of samples between $1000$ and $20$, for 3-SAT at $\alpha=3.0$
we report a scaling of $N^{1.23}$ whereas for 4-SAT at $\alpha=8.0$
we found a scaling of $N^{1.18}$.

\end{document}